\documentclass[12pt]
{revtex4}

\usepackage{latexsym}
\usepackage{amssymb}
\usepackage{graphicx}
\usepackage{dcolumn}
\usepackage{bm}
\input{epsf}

\begin{document}

\title{Discussing Cosmic String Configurations in a Supersymmetric Scenario without Lorentz Invariance}

\author{C. N. Ferreira}

\email{crisnfer@pq.cnpq.br}

\affiliation{
N\'ucleo de Estudos em F\'{\i}sica, 
Instituto Federal de Educa\c{c}\~ao, Ci\^encia e Tecnologia Fluminense \\
        Rua Dr. Siqueira, 273, Campos dos Goytacazes,\\
         Rio de Janeiro, Brazil, CEP 28030-130\\}

\author{J. A. Helay\"el-Neto}

\email{helayel@pq.cnpq.br}

\affiliation{
Centro Brasileiro de Pesquisas F\'{\i}sicas, Rua Dr. Xavier Sigaud 150,
Urca,\\ Rio de Janeiro, Brazil, CEP 22290-180}

\author{C. E. C. Lima }

\email{ceclima@ig.com.br}
	
\affiliation{
Centro Brasileiro de Pesquisas F\'{\i}sicas, Rua Dr. Xavier Sigaud 150,
Urca,\\ Rio de Janeiro, Brazil, CEP 22290-180}


\begin{abstract}
The main goal of this work is to pursue an investigation of cosmic string configurations  focusing on possible 
consequences of the Lorentz-symmetry breaking by a background vector. We 
analyze the possibility of cosmic strings as a viable source for fermionic Cold Dark Matter particles. 
Whenever the latter are charged and have mass of the order of $10^{13}GeV$, we propose they could decay into usual cosmic rays. 
We have also contemplated the sector of neutral particles generated in our model. Indeed, being neutral, these particles are hard to be detected; however, by virtue of the Lorentz-symmetry breaking background vector, it is possible that they may present an 
electromagnetic interaction with a significant magnetic moment.

\end{abstract}



\maketitle


\section{Introduction}

The mechanisms for breaking symmetries in theories of fundamental Physics may yield many interesting spectral relations among particles. In Quantum Field Theory, important invariances appear in connection with the Standard Model for Particle Physics. These symmetries are the Lorentz, CPT\cite{CPT} and Supersymmetry (SUSY) \cite{Weinberg} invariances. Lorentz- and CPT-symmetry breakings\cite{Colloday98,Kostelecky89} can occur if we consider processes at energy scales close to the Planck mass \cite{Carroll:1989vb,Kostelecky99}. 
In Quantum Electrodynamics, there is a number of works that present constraints on the Lorentz-symmetry violating parameters \cite{Jacobson:2003bn,Stecker:1992wi}. One of the most important processes for which the breaking of Lorentz symmetry may be measurable is related to high-energy $\gamma $-rays from extragalactic sources. The idea in the works that treat this question is that the $\gamma $-rays are absorbed during their interaction with the low-energy photons of intergalactic radiation \cite{Stecker:1992wi,Stecker:1998ib}; there occurs annihilation into electron-positron pairs in intergalactic space. This sort of physical process imposes constraints on the Lorentz-invariance violation parameters, which can also yield bounds from the astrophysical tests\cite{Stecker:1992wi,Stecker:2004xm,Stecker:2004vm}. 

To study these phenomena, some authors proposed alternative models that can describe Lorentz-symmetry breakings and their effects\cite{Kostelecky89,Carroll:1989vb,Kostelecky95}. 
The other symmetry we study in our work is supersymmetry (SUSY), which is the main ingredient of new theories, such as the Physics Beyond the Standard Model; it appears as a viable framework that solves some problems of the Standard Model and gives us possible candidates to Dark Matter (DM). The astronomical
evidence for additional, non-luminous matter, or dark matter, strengthens our motivations to consider SUSY. Another motivation to adopt SUSY  is the fact that this symmetry appears as the  main ingredient of String Theory\cite{Green87}.  In the Minimally Supersymmetric Standard Model (MSSM), the breaking of SUSY is soft, which is an attractive way of solving the hierarchy problem and linking the electroweak scale to physics close to the Planck scale. This soft SUSY breaking can be related with the Lorentz-invariance violation (LIV) in the study of the vortex superfluids\cite{Ferreira:2008zz}, where the LIV is realized by a Kalb-Ramond field\cite{Davis:1989gn}.
 
The Kalb-Ramond and dilaton fields appear in heterotic string theory and can couple to the Yang-Mills-Maxwell-Chern-Simons field as the result of a quantum effect. 
The Kalb-Ramond field can play the role of the background field in \cite{Majumdar:1999jd,Maity:2004he}. In another context, vortex configurations may  be studied in the brane-world framework \cite{Braga:2004ns} and present important results in a supersymmetric scenario \cite{Morris:1995wd,Davis:1997bs,Ferreira:2000pi,Ferreira:2002mg,NunesFerreira:2005if}. These motivations are the basis for this work, where we consider a rather general set of background fields that induce LIV in connection with possible supersymmetric cosmic string configurations. Structures like cosmic strings \cite{Vilenkin:1981zs,Hiscock:1985uc,Gott:1984ef,Garfinkle:1985hr,Hindmarsh:1994re,Vilenkin1,Kibble:1980mv}, probably produced during phase transitions \cite{Kibble:1976sj}, appear in some
Grand-Unified gauge theories and carry a huge energy density \cite{Hindmarsh:1994re}. They have been studied to provide a possible origin for the seed density perturbations which became the large
scale structure of the Universe observed today \cite{Sato:1986pd}. These fluctuations would
leave their imprint in the Cosmic Microwave Background Radiation (CMBR) that
would act as seeds for structure formation and, then, as builders of the
large-scale structures in the Universe \cite{Turok:1985tt}. However, they have presently been discarded as the only responsible for structure formation. They have re-acquired renewed phenomenological interest in connection with  String Theory \cite{Kibble:2004hq,Polchinski:2004ia}. They may also help to explain the most energetic events in the Universe, such as ultra-high-energy cosmic rays\cite{Bhattacharjee:1989vu,Bhattacharjee:1991zm,MacGibbon:1989kk} above the Greisen-Zatsepin-Kusmin (GZK) cut-off, that lies on energies of the order of 5 x $10^{10}$ GeV \cite{Kuzmin:1998kk}.
In this work, we analyze possible consequences  of supersymmetric cosmic strings interacting\cite{Bezerra:2004qv,Ferreira:2002mg} with a vector background responsible for the Lorentz breaking effect \cite{Belich:2003fa}. This scenario may yield consequences to particle creation and this is an issue we devote special attention to in the present paper. 

The outline of this work is as follows: in Section II, we start by presenting a simple model with a Lorentz-symmetry breaking contribution formulated in terms of superfields. In Section III, we study vortex configurations and we discuss the physical implications of this Lorentz-symmetry violation.
We also analyze the potential, the magnetic flux and the currents induced by LIV. In Section IV, we compute the propagator for the excitations in the gauge sector of the model and comment on the  role  of the Lorentz-symmetry breaking additional terms. Section V is devoted to  the study of the fermionic charged fields that  can be ejected from cosmic strings. We consider the possibility that 
these fields may be interpreted as the responsible for charged Cold Dark Matter (CDM) particles. In Section VI, we set up the part of the model that accounts for the neutral particles that may appear as (neutral) CDM particles. We give here special room to the discussion of the possible magnetic moment interactions of these neutral particles. Finally, in Section VII, we draw our General Conclusions with a discussion on phenomenological aspects of our model.
 
\section{Supersymmetric Cosmic string model with Lorentz-Invariance Violation}

In this Section, we present the general features of a supersymmetric model that may produce cosmic string configurations with LIV. To accommodate 
cosmic strings in a supersymmetric context, we need a family of chiral scalar superfields, $\Phi_i (\phi_i , \xi_i , G_i) $. These superfields contain the complex physical scalar fields, $\phi_i $, their fermionic partners, $ \xi_i $, and complex auxiliary fields, $G_i $; they can be written as $\theta$-expansions according to the parameterization below:  
\begin{equation}
\Phi_i = e^{-i\theta \sigma^{\mu} \bar \theta \partial_{\mu}}[\phi_i(x) +
\theta^{a}\xi_{i a}(x) + \theta^2G_i(x)]\, , \label{sup1}
\end{equation}
where the label $i$ represents the flavours of the superfields needed to the correct description of the cosmic string. 
The other fundamental ingredient for a cosmic string configuration is the gauge field, $H_{\mu}$. 
To accommodate this field, we consider a gauge superfield, ${\cal V}_{_{\!H}} $, which also contains a fermionic gauge partner, $\chi $, and a real auxiliary field, $D $. The superfield ${\cal V}_{_{\! H}}$ is already assumed to be in the Wess-Zumino gauge, with a $ \theta $- expansion as follows:
\begin{equation}
{\cal V}_{_{\! H}} = \theta \sigma^{\mu} \bar \theta H_{\mu}(x) + \theta^2 \bar \theta
\bar \chi (x) + \bar\theta^2 \theta \chi (x)+ \theta \bar \theta^2
D(x)\, . \label{sup2}
\end{equation}
The  cosmic string Lagrangian, ${\cal L}_{_{CS}}$,  that is $U(1)$-gauge symmetry invariant and presents $N=1$ supersymmetry is
\begin{equation}
{\cal L}_{_{CS}} = \bar \Phi_i \,e^{^{\!2 q_i\, {\cal V}}}\! \!\Phi_i |_{_{\theta \theta \bar \theta \bar \theta}} \! \! + \alpha \,
 {\cal H}^{\alpha} \, {\cal H}_{\alpha} |_{_{\theta \theta}}\! \! \, +W|_{_{\theta \theta}} + h.c., \label{acaoqed}
\end{equation}
where $\alpha $ is a real parameter, h.c. is the Hermitian conjugate, and the field-strength superfield, $ {\cal H}^a $, is  written as 
${\cal H}^a = -\frac{1}{4} \bar {\cal D}^2 {\cal D}^a {\cal V}_{_{H}}$. The supersymmetry covariant derivatives are given as ${\cal D}_a = \partial_a - i \sigma^{\mu}_{a \dot a} \bar \theta^{\dot a}
\partial_{\mu} $ and $ \bar {\cal D}_{\dot a} =- \partial_{\dot a} + i \theta^a \sigma^{\mu}_{\dot a a}
\partial_{\mu}, $ where the $\sigma^{\mu}$-matrices read as $\sigma^{\mu} \equiv ({\bf 1}; \sigma^i)$, the $\sigma^i$'s
being the Pauli matrices.

To realize the U(1)-gauge symmetry breaking responsible for the cosmic string vacuum, we consider a superpotential, $W$,
as given by
\begin{equation}
W = \beta\Phi_0\Big(\Phi_+\Phi_- - \eta^2\Big),\label{supot}
\end{equation}
where $\eta $ and $\beta$ are real parameters. Then, we consider in (\ref{acaoqed}),$\, \, i = 0,\, +, \, - $ \, with $q_0= 0 $ related to the neutral superfield, $q_+ =1$ associated to the chiral supermultiplet and $q_- = -1 $ to the anti-chiral supermultiplet.

In a scenario with SUSY, this structure is  sufficient to obtain a cosmic string \cite{Morris:1995wd}; but, in our case, we wish to study this phenomenon in connection with LIV. This fact changes the cosmic string configuration giving us new features.

Models with LIV without SUSY have been already studied\cite{Bezerra:2004qv}; nevertheless, if we consider events that occur in the early universe, as the cosmic string formation, we need to include  SUSY. In this analysis, we consider the cosmic string in the presence of the supersymmetric version of the Maxwell-Chern-Simons term; so, we need another gauge superfield, $ {\cal V}_{_{\! A}}$, that has the same form as the one given in eq.(\ref{sup2}) and is also taken in the Wess-Zumino gauge:

\begin{equation}
{\cal V}_{_{\! A}} = \theta \sigma^{\mu} \bar \theta A_{\mu}(x) + \theta^2 \bar \theta
\bar \lambda (x) + \bar\theta^2 \theta \lambda (x)+ \theta \bar \theta^2
\tilde D(x)\, . \label{sup3}
\end{equation}
Also, we have to add a dimensionless chiral superfield, $\Omega(\omega, \psi,I)$, given by  
\begin{equation}
\Omega = e^{-i\theta \sigma^{\mu} \bar \theta \partial_{\mu}}[\omega(x) +
\theta^{a}\psi_a(x) + \theta^2I(x)],  \label{sup4}
\end{equation}
which carries a complex scalar field, $\omega $, a fermionic partner, $ \psi$, and an auxiliary field, $ I$. The Lagrangian that accommodates the LIV term reads as below
\begin{equation}
{\cal L}_{_{MCS}} =  {\cal F}^a{\cal F}_a|_{_{\theta \theta }}  + {\cal H}^a{\cal F}_a\Omega|_{_{\theta \theta }} + h. c. \, .\label{MCS}
\end{equation}

The Lagrangian ${\cal L}_{_{MCS}}$ is written in terms of a field-strength superfield, ${\cal F}^a= -\frac{1}{4} \bar {\cal D}^2 {\cal D}^a {\cal V}_{_{A}} $, that accommodates the electromagnetic field, $A_{\mu}$, the photino, $\lambda$, and a real auxiliary field, $ \tilde D $. 

For a detailed description of the component fields and degrees of freedom displayed in (\ref{sup3}), the
reader is referred to the paper of ref. \cite{Belich:2003fa}. As it can be readily checked, this action is invariant under Abelian gauge and SUSY transformations \cite{Ferreira:2002mg}. The SUSY transformations act upon the string bosonic configuration, and the fermionic zero-mode excitations turn out to be:
\begin{equation}
\begin{array}{ll}
\xi^{\pm}_a = i \sqrt{2}\sigma^\mu_{a
\dot a}\bar \varepsilon ^{\dot a}D_{\mu} \phi^{\pm} + \sqrt{2} \varepsilon_a G^{\pm},\\
\\
\xi^{0}_a = i \sqrt{2}\sigma^\mu_{a
\dot a}\bar \varepsilon ^{\dot a}\partial_{\mu} \phi^{0} + \sqrt{2} \varepsilon_a G^{0},
\\
\\
\chi _a = \sigma^{\mu \nu}_{a\dot a}\varepsilon^{\dot a}H_{\mu \nu}
+ i \varepsilon_a D ,\\
\\
\psi_a = i \sqrt{2}\sigma^{\mu}_{a \dot a} \bar \varepsilon ^{\dot a} \partial_{\mu} \omega , 
\\
\\
\lambda_{a} = \sigma^{\mu \nu}_{a \dot a} \varepsilon^{\dot a} F_{\mu \nu} + i \varepsilon_a \, \tilde D .
\end{array}
\label{ferm2}
\end{equation}

The physical interpretation of the Lagrangian ${\cal L}_{MCS} $ is that the dimensionless superfield, $ \Omega$, represents the background that can give us information about the early universe when cosmic strings were formed. In this context, LIV can become an important ingredient for cosmic string appearance. This cosmic string is superconducting and has a different characteristic with respect to the usual and Witten's cosmic strings \cite{Witten:1986qx}.

\section{Cosmic String Interactions in a Framework with LIV}

In this Section, let us analyze cosmic string configurations. We study the equations of the motion and show that it is possible to find solutions in our model. 
In components, we have a $U(1) \times U(1)'$ mixing-term \cite{Foot:1994vd}, which can be diagonalised by considering the following field reshufflings:

\begin{eqnarray}
H_{\mu} &\rightarrow & (1-\rho^2)^{-1/2} H_{\mu}\\
A_{\mu} &   \rightarrow& A_{\mu} -\rho(1-\rho^2)^{-1/2}H_{\mu}, \label{gaugebasis}
\end{eqnarray}
where we write the real part of $\omega$ as $(\omega + \omega^*) \equiv \rho$. The mixing - term comes from the Lagrangian of equation (\ref{MCS}). 

In this work, we choose the condition $\alpha +2 \rho =1$ to specify the  background field configurations for the $\omega$-field, in such a way that we have $\rho = {\rm constant}$. The constant $\rho$ is, in our analysis, a physical parameter 
and cannot be completely scaled away in the presence of the Lorentz-breaking interaction. In string theory \cite{Green87}, the $\rho $-field may exhibit interaction with the gauge potential; here, with a LIV environment, it freezes at a constant value in the background.

For the on-shell version of the present model, we have the bosonic Lagrangian:

\begin{eqnarray}
L_B=  2 D_{\mu} \phi [D^{\mu} \phi]^* - \frac{1}{4}H_{\mu \nu} H^{\mu \nu} -\frac{1}{4} F_{\mu \nu} F^{\mu \nu}\nonumber \\
\! \! \!+ {1 \over 2} (1\!- \!\rho^2)^{^{\!\!-1\!/2}} \! \epsilon^{^{\mu \alpha \beta \nu}}\! \! \!\! \! \! \! v_{\mu}\, H_{\alpha \beta}\Big[\!\rho (1-\rho^{2})^{^{\!\!-1/2}}H_{\nu} \!+ \!A_{\nu}\!\Big]\!-\! U,
\label{lag4}
\end{eqnarray}
where the Lorentz-symmetry breaking term that contains the antisymmetric part of $\omega$ is written as $(\omega - \omega^*) = i\Delta$.
The form of $\Delta $ is fixed in terms of the LIV parameter, since we define $v_{\mu} = i\partial_{\mu} \Delta $ as a constant.  This ansatz for $\rho$ and $v_{\mu}$ is possible whenever we consider a static background. 

The bosonic fields, in polar coordinates, may be parameterized as (axially symmetric ansatz):
\begin{eqnarray}
\phi_0 & = & 0 \\
\phi_{_+} &= &\phi_-^* = f(r) e^{in\theta}  = \phi\\
H_{\mu} &=& H_{\theta}(r)\delta_{\mu}^{\theta}\\
G_{\pm} &= & D =0 \\
G_0 & = &\mu \Big(\eta^2 - f(r)^2 \Big)\, ,\label{ansatz1}
\end{eqnarray}
where $ G_{\pm}$, $D$ and $G_0$ are the auxiliary fields introduced in the previous section.

We impose as boundary conditions:
\begin{equation}
\begin{array}{lllll}
f(r) =0 &  r=0 & & f(r) = \eta  & r \rightarrow \infty .\\
\end{array}
\end{equation}

The gauge covariant derivative is:
\begin{equation}
D_{\mu} \phi = [\partial_{\mu} + i e(1-\rho^2)^{-1/2}H_{\mu}] \phi .
\end{equation}

The potential turns out to be given by: 
\begin{eqnarray}
U &= &  \beta(\bar \phi_+ \bar \phi_- - \eta^2)(\phi_+\phi_- - \eta^2  ) \nonumber\\
&+&  \beta \bar \phi_0 \phi_0\Big(\bar \phi_+ \phi_+ + \bar \phi_- \phi_- \Big) \, .\label{lag1}
\end{eqnarray}
With the ansatz (\ref{ansatz1}) for the cosmic string, $U $  can be split according to:
\begin{equation}
U = \beta \, \Big( |\phi|^2 - \eta^2\Big)^2.
\label{potcs}
\end{equation}

Now, let us analyze the charge induced by the Lorentz-symmetry breaking in the cosmic string core. 
The field equations  imply that $ \phi $ and $H_{\mu}$ are solutions to the following 
differential equation:

\begin{equation}
\partial_{\mu} \partial^{\mu} \phi +e^2 (1-\rho^2)^{-1}H_{\mu} H^{\mu} -{\partial V(\phi) \over \partial \phi} =0,
\end{equation}
which yields,

\begin{equation}
{1 \over r} {d \over dr}\Big(r {d f\over dr}\Big) -\Big[\Big({n \over r} -e(1 -\rho^2)^{-1/2}H\Big)^2 - 2\beta(f^2-\eta^2)\Big] f=0.
\end{equation}

The equations for the gauge fields are given by

\begin{equation}
\partial_{\mu} H^{\mu \nu} = {\cal J}^{\nu} + {\cal J}_{top}^{\nu},
\end{equation}
where  the 4-dimensional current density, $ {\cal J}_{\nu}$, is related to the cosmic string fields according to
\begin{equation}
{\cal J}_{\mu} = - ie \left[ \phi^* \partial_{\mu } \phi - \phi
(\partial_{\mu} \phi)^*\right] + 2e (1-\rho^2)^{-1} |\phi |^2 H_{\mu}.
\label{curr2}
\end{equation}

Let us consider the Lorentz-symmetry violating vector, $v_{\mu}=(0,0,0,v)$, 
being so chosen that, whenever $\mu =z$, it is connected with the charge, 
$Q$. The $ {\cal J}_{top}^{\mu}$ is the topological current,  with the 
extra gauge field, $ A_{\mu}$, given by

\begin{equation}
{\cal J}_{top}^{\mu} = -(1-\rho^2)^{-1/2}\epsilon^{\mu \nu \alpha \beta}v_{\nu} F_{\alpha \beta} .
\end{equation}

The equation for the gauge potential along the angular  component is

\begin{equation}
B_z = {1 \over r}{d \over dr}\Big( r H(r)\Big)
\end{equation}

\begin{eqnarray}
{dB_z \over dr} + 2 e(1-\rho^2)^{-1/2}& \Big[{n \over r} + e (1-\rho^2)^{-1/2}H\Big]f^2 \, + \nonumber \\ v(1-\rho^2)^{-1/2} E=0\, , &
\end{eqnarray}
where the magnetic field is defined as $ B_k = \epsilon_{i j k}H^{i j} $ and the electric field is given by $F_{0r} = -E$. The equation for the gauge field, $A_{\mu}$, is given by

\begin{equation}
\partial_{\mu}F^{\mu \nu} = (1-\rho^2)^{-1/2}\epsilon^{\nu \kappa \alpha \beta} v_\kappa H_{\alpha \beta};
\end{equation}
In terms of the magnetic and electric fields, we have

\begin{equation}
{1 \over r}{d \over dr} \Big(r \, E(r) \Big)  = v (1-\rho^2)^{-1/2} B_z(r) .  
\end{equation}

\begin{figure}
\centering
\includegraphics[width=8cm]{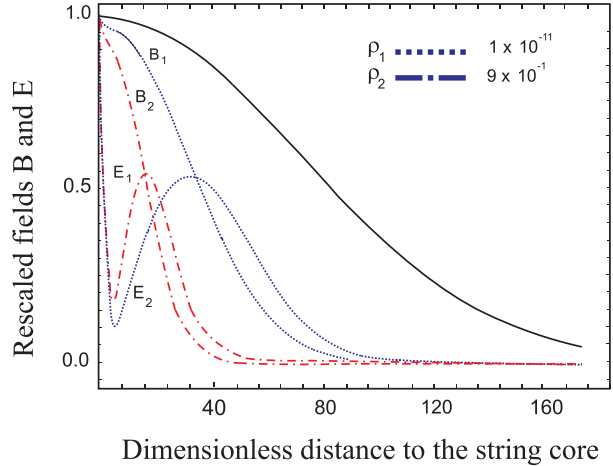}
\parbox{5in}{\caption{\ In this graph, we plot the behavior of the electric  and \, magnetic \,  fields \, of \, the\,  cosmic \, string with mixing parameter, $\rho$. We consider $v =0.03$.
 }}\label{Figura1} 
\end{figure}

The magnetic flux is quantized as below:
\begin{eqnarray}
\phi_{Mag} & = &\oint H_i rdx^i = - 2 \frac{\pi}{e} n (1-\rho^2)^{1/2}.\label{flux}
\end{eqnarray}

\begin{figure}
\centering
\includegraphics[width=8cm]{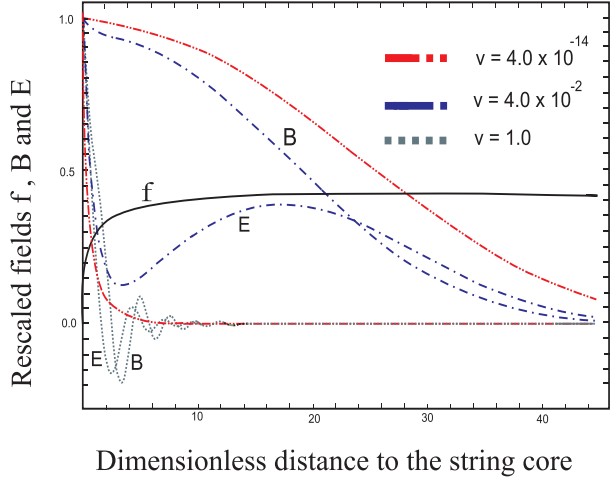}
\parbox{5in}{\caption{ In this graph, the point curve shows the behavior  of the cosmic string with parameter\, $v$ . The\, solid \, curve \, is the ordinary cosmic string solution, i.e.\, $v=0$ \, and \, $\rho=0$.  The parameter of the other curves is $ \rho = 10^{-11}$.
 }}\label{Figura2} 
\end{figure}

The region of validity for  $\rho$ ( $\rho <1 $, actually, $\rho $ is a very small parameter) ensures that the magnetic flux is non-vanishing and conserved, as depicted in Figure 1, in terms of the variation of the parameter $ \rho$.  

The solution to this problem is the change of the cosmic string configuration to include $A_t \neq 0$ in the $v_z$ case.  In Figure 2, we plot the solution of the electric field, E, magnetic field, B, and scalar, f, in terms of the dimensionless radial coordinate. This graph is important to analyze the convergence of the fields, which illustrates the stability of our cosmic string solution. The comparison between  our solutions and the ordinary cosmic string solution shows that an electric field appears and  falls off  as $1/r $, when $r \rightarrow \infty $. This is compatible with the vortex configuration. 

It is relevant to analyze the fact that this cosmic string is charged and the symmetry whose breaking is involved here is Lorentz invariance, different from the Witten's superconducting cosmic string. In the Witten's framework, the superconductivity is given by the breaking of the $U(1) $ gauge symmetry in the core, which gives the current and the preserved gauge symmetry in the vacuum that responds for the long-range behavior of the electromagnetic field, $F_{\mu \nu} $. This mechanism includes two complex scalar fields that interact through a more complicated potential. In our approach, the unique gauge symmetry that is broken is the U(1)-group concerned to the cosmic string configuration; for this reason, the potential includes only the cosmic string field, $ \phi$. 

\section{The Cosmic String Propagator}

Another important ingredient that we have to analyze are the 
propagator of the gauge-field sector. There is  a mixing of the  $A_\mu$-and $H_\mu$ -potentials given by the LIV term. The poles of the propagator and their corresponding residues allow us to infer about the spectrum of spin-0 (longitudinal) and spin-1 (transverse) excitations: we have to be sure that neither tachyonic poles nor ghost-like states are present in the model. Specially now, that both $A_\mu$ and $H_\mu$ are mixed and an external background vector, $v^\mu$, appears that may, in general, yield massive poles, we must guarantee that the gauge-field sector is not plagued by unphysical modes.

With this purpose, we parameterize $ \phi $ as $\phi =[{ \tilde \phi} (x)^{\prime }+\eta ]e^{i\Sigma (x)}$,
where $ \tilde \phi ^{\prime } $ is the quantum fluctuation around the ground state, $\eta$. 
We concentrate on the bosonic Lagrangian in terms of the physical fields, $\tilde \phi ^{\prime } $, 
$H_\mu $ and $A_{\mu}$,  and we adopt the unitary gauge for the broken U(1)'-factor (associated 
to $H_{\mu}$). This gauge choice $ \Sigma =0 $ is in perfect agreement with the Wess-Zumino gauge 
adopted in the eq. (\ref{sup2}) of Section 2. By fixing this gauge, we remove away compensating 
fields introduced by SUSY. We are still left with the usual gauge freedom, so that we still have 
the freedom to fix the unitary gauge. For the sake of reading off these propagator, we refer to 
the bosonic Lagrangian (\ref{lag4}). We first write it in a more convenient form:

\begin{equation}
{\cal L} = \frac{1}{2}\sum_{\alpha \beta} {\cal A}^{\alpha} {\cal O}_{\alpha
\beta} {\cal A}^{\beta},
\end{equation}

\noindent
where ${\cal A} ^\alpha =(A^\mu ,H^{\mu }, A^{\mu})$ and ${\cal O}_{\alpha \beta }$ is the wave operator. We notice that $\Sigma$ mixes with $A^{\mu}$. However,  we adopt the t'Hooft $R_{\xi}$-gauge and they decouple from each other. So, the $\Sigma-\Sigma$ propagator can be derived independently from the propagator for the ($A^{\mu} ,H^{\mu }$) sector. We apply the usual procedure to invert the operator $ {\cal O}^{-1}_{\alpha
\beta} $ in order to find the gauge-field propagator of this problem.

To read off the gauge-field propagator, we shall use an extension of the spin-projection operator formalism presented in \cite{Kuhfuss:1986rb}. The new aspect in this work is that, describing the LIV terms in connection with the cosmic string fields, we have to add other new operators coming from the Lorentz-symmetry breaking terms and the cosmic string interactions. Then, we need the usual two operators, $\Theta_{\mu \nu}$ and $\Gamma_{\mu \nu}$, being respectively, the transverse and longitudinal projection operators, given by $
\Theta _{\mu \nu }=\eta _{\mu \nu }-\Gamma _{\mu \nu }$,
and $\Gamma_{\mu \nu} = \frac{\partial_{\mu} \partial_{\nu}}{\Box }$. 
In order to find the inverse of the wave operator, let us calculate the products of operators for all non-trivial combinations involving the projectors. The relevant multiplication rules are listed in Table I, where the new spin operator coming from the Lorentz breaking sector is   $\Omega_{\mu \nu}$, defined in terms of Levi-Civita tensor as

\begin{table}
\caption{Multiplicative table for spin operators}
\begin{center}
\begin{tabular}{|c|c|c|c|c|c|c|}\hline
&$\theta^{\alpha}_{\hspace{.3 true cm} \nu}$  & $\Gamma^{\alpha}_{\hspace{.3 true cm} \nu}$ &$\Omega^{\alpha}_{\hspace{.3 true cm} \nu}$& $\Lambda^{\alpha}_{\hspace{.3 true cm} \nu} $&
$\Sigma^{\alpha}_{\hspace{.3 true cm} \nu} $& $\Sigma^{\hspace{.3 true cm} \alpha}_{\nu} $\\
\hline
$\theta_{\mu \alpha} $& $\theta_{\mu \nu}$ & 0 & $\Omega_{\mu \nu} $& $\Lambda_{\mu \nu} - \frac{\lambda}{\Box} \Sigma_{\mu \nu}$& $\Sigma_{\mu \nu} - \lambda \Gamma_{\mu \nu}$&0\\
\hline
$\Gamma_{\mu \alpha }$&0 & $\Gamma_{\mu \nu} $& 0 & $ \frac{\lambda}{\Box} \Sigma_{\nu \mu}$& $\lambda \Gamma_{\mu \nu} $ & $ \Sigma_{\nu \mu}$\\
\hline
$ \Omega_{\mu \alpha} $& $ \Omega_{\mu \nu}$ & 0 & $f_{\mu \nu}$ & 0 & 0& 0\\
\hline
$\Lambda_{\mu \alpha} $& $ \Lambda_{\mu \nu} - \frac{\lambda}{\Box} \Gamma_{\mu \nu}$& $\frac{\lambda}{\Box} \Sigma_{\mu \nu}$& 0 & $v^2 \Lambda_{\mu \nu}$ & $v^2\Sigma_{\mu \nu}$&
$\lambda \Lambda_{\mu \nu}  $\\
\hline
$ \Sigma_{\mu \alpha }$ & 0 &$ \Sigma_{\mu \nu}$ & 0 & $\lambda \Delta_{\mu \nu} $ & $\lambda \Sigma_{\mu \nu} $&  $ \Lambda_{\mu \nu}\Box $ \\
\hline
$\Sigma_{\alpha \mu } $ & $\Sigma_{\nu \mu} - \lambda \Gamma_{\mu \nu}$ & $ \lambda \Gamma_{\mu \nu} $ & 0 & $v^2 \Sigma_{\nu \mu} $ & $v^2 \Box \Gamma_{\mu \nu}$ & $\lambda \Sigma_{\nu \mu }$\\
\hline
\end{tabular}
\end{center}
\end{table}

\begin{equation}
\Omega_{\mu \nu} = \epsilon_{\alpha \beta \mu \nu} v^{
\alpha} \partial^{\beta}, 
\end{equation}
and the operator $ f_{\mu \nu}$, that we find by squaring $\Omega $, gives us 

\begin{equation}
f_{\mu \nu} \equiv  \Omega_{\mu \alpha} \Omega^{\alpha}_{  \nu } = M_7 \Gamma_{\mu \nu} + M_8 \Lambda_{\mu \nu} + M_9 \Sigma_{\mu \nu},
\end{equation}
so that we have to define other operators such as

\begin{eqnarray}
\Sigma_{\mu \nu} &=& v_{\mu } \partial_{\nu}, \\
\lambda \equiv \Sigma_{\mu }^{\mu} &= & v_{\mu }\partial^{\mu}, \label{30}\\
\Lambda_{\mu \nu} &= & v_{\mu} v_{\nu}.
\end{eqnarray}

The operators $\Sigma_{\mu \nu} $ and $ \lambda $ project the longitudinal 
part of a vector field along the $v^{\mu} $-direction, while $ \Lambda_{\mu \nu}$
projects the whole vector field
(longitudinal plus transverse part) along $ v^{\mu}$.

We write below the explicit expressions for the propagator we are interested in:

\begin{eqnarray}
<AA> &=& \frac{i}{M_1} \Theta_{\mu \nu} +
\frac{i}{(M_2 -M_3^2M_4^{-1} M_7) }
\Gamma_{\mu \nu} \nonumber \\
&-& \frac{i}{ M_3^2 M_4^{-1} M_8} \Lambda_{\mu \nu} -   \frac{i}{M_3^2 M_4^{-1} M_9} \Sigma_{\mu \nu} \;  \label{propag1}
\end{eqnarray}

\begin{eqnarray}
<HH> &=& \frac{i}{M_4} \Theta_{\mu \nu} +
\frac{i}{(M_5 -M_3^2M_1^{-1} M_7) }
\Gamma_{\mu \nu} \nonumber \\
&-& \frac{i}{ M_3^2 M_1^{-1} M_8} \Lambda_{\mu \nu} -   \frac{i}{M_3^2 M_1^{-1} M_9} \Sigma_{\mu \nu} \;
,
\label{propag2}
\end{eqnarray}

\noindent
where the first term in both equations is the transverse part of the propagator, the second is the longitudinal part, the other two terms are related with the LIV, and $M_1,...,M_9$ are quantities that read as the following expressions:

\begin{equation}
\begin{array}{lll}
M_1 = M_8 =  \Box , \hspace{.3 true cm}  M_2=  - \frac{1}{\epsilon}\Box , \hspace{.3 true cm} M_3 = M_6= - {1\over (1-\rho^2)^{1/2} },\\
\\
M_4=  \frac{1}{2}\Box + 2e^2(1-\rho^2)\eta^2, \hspace{.5 true cm} M_5= 2e^2(1-\rho^2) \eta^2 ,\\
\\
M_7 = [-v^2 \Box - \lambda^2], \hspace{.5 true cm} M_9 =  \frac{\lambda}{2},
\end{array}
\end{equation}
where $ \lambda $ is  the momentum space expression for the operator (\ref{30}).

In our notation, $<AA>$  represents the propagator for the gauge field $A_{\mu}$, and $<HH>$ stands for the propagator between the gauge fields $H_{\mu}$ that is connected with the cosmic string. 

By  analyzing the $<AA>$ propagator, we can show that the transverse part is given by $M_1 $. It has a trivial pole, then this field describes a massless excitation and its $U(1)$- symmetry does not break down. This analysis is important because it gives us  information on the range of the field. 
The fact that the field $A_{\mu}$ presents an exact $U(1)$ gauge symmetry gives us 
the interpretation that is associated to a long-range interaction. 
This propagator displays a pole at $k^2 = 0 $ also in the $ \Lambda_{\mu \nu}$-sector, which indicates 
the presence of a massive excitation along the background $ v^{\mu}$.
This is not the case for the $ \Lambda_{\mu \nu}$-sector of the  $<HH>$ propagator. It is interesting to 
notice that  the $<AA> $ - and - $<HH>$ propagator present a pole of the type $ v^{\mu}k_{\mu}=0$, which
points to  an excitation that propagates  orthogonally to $ v^{\mu}$.

As for the propagator (\ref{propag2}), the transverse part presents a mass given by $4e^2(1-\rho^2)\eta^2 $.
We notice that this mass presents a parameter $ \rho$, connected with the LIV. This parameter is very important for the understanding of the range of the gauge field $ H_{\mu}$. If $\rho =1$, the field is free, but the vortex disappears because, in this limit, we have, by eq. (\ref{flux}), a zero magnetic flux. Then, the parameter $\rho$ is important to give us a fine-tuning of the interaction.

\section{Fermionic Charged Particles from Cosmic String configuration}

In this Section, we shall consider the sector of fermionic excitations that propagate on the background settled down by the bosonic cosmic string configuration.
Let us consider the Lagrangian for the $\Xi $-field:

\begin{equation}
{\cal L}_{CC} = i \bar \Xi_+ \gamma^{\mu} D_{\mu} \Xi_+ +i \bar \Xi_- \gamma^{\mu} D_{\mu} \Xi_- + i \bar \Xi_0 \gamma^{\mu} \partial_{\mu} \Xi_0, \label{Fer2}
\end{equation}
where $ {\cal L}_{CC} $ is the Lagrangian piece that contains the charged fermionic fields written down in terms of four-component Majorana spinors as follows:

\begin{equation}
\Xi_{\pm} \equiv \Xi_1 \pm i \Xi_2 
\end{equation} 
where $ \Xi_1$ and $ \Xi_2$ have the form $\Xi(x) = \Big(\begin{array}{ll}\xi_a(x)\\
\bar \xi^{\dot a}(x) \end{array}\Big)$.
The gauge covariant derivatives are:

\begin{eqnarray}
D_{\mu} \Xi_+ &=& (\partial_{\mu} + i e(1-\rho^2)^{-1/2}H_{\mu}) \Xi_+, \nonumber\\
D_{\mu} \Xi_- &=& (\partial_{\mu} - i e(1-\rho^2)^{-1/2}H_{\mu}) \Xi_- \, .
\end{eqnarray}
The Yukawa Lagrangian is given as below:

\begin{equation}
\begin{array}{ll}
{\cal L}_Y = &  \beta \, \, \Big(\phi_- \Xi_0 \Xi_+ +\phi_+ \Xi_0 \Xi_- \Big) + h. c..
\end{array}
\end{equation}
We propose that the fermionic solution, in the inner region of the cosmic string, displays, in cylindric coordinates, the form:
\begin{equation}
\Xi_i = \Xi(r,\theta)_i e^{\alpha_i(z-t)},
\end{equation}
where $\alpha_i(z-t)$ represents the left-moving superconducting currents flowing along the string at the speed of light, because in the core of the cosmic string, the fermions do not have mass, because $\phi_i =0$. In this case, the Lagrangian (\ref{Fer2}) gives us the zero-modes inside the string.

The other important aspect are the currents inside the string: they are conserved currents. One of them is the Noether current, given by  
\begin{eqnarray}
{\cal J}_{_{(\Xi)}}^{a} &=& -{e \over (1-\rho^2)^{1/2}} \Big[\bar \Xi_+ \gamma^{a} \Xi_+- \bar \Xi_- \gamma^{a} \Xi_- \Big].\label{curr1}
\end{eqnarray}
This current only involves the fermionic SUSY partners of the cosmic string scalar fields. Inside the string,  SUSY is broken in the range of $10^{11} - 10^{13}$ GeV and there appears a fermionic current. This current, described by (\ref{curr1}), can be in the $z$-direction, carrying an electric field, $E_z $. The dynamics of the current can be given by

\begin{equation}
{d J_z \over dt} = {e^2\over (1-\rho^2)} E_z \, .
\end{equation}

We notice that the variation of the current has a dependence on the LIV parameter $\rho $. This current grows until produced particles are ejected from the cosmic string. These charged fermions particles acquire masses 
from the breaking of the U(1) gauge symmetry where the scalar cosmic string field $<|\phi_{\pm}|> $ has a vacuum value,  $\eta $.  The fermionic particle masses studied here are described by Yukawa's term that reads after a gauge symmetry breaking
\begin{equation}
{\cal L}_Y =   \beta \, \, \eta \, \Big(\Xi_0 \Xi_+ + \Xi_0 \Xi_- \Big). \label{mass2}
\end{equation}      

The physical interpretation of these particles could be formulated in the context ultra-energetic cosmic rays, above the Greisen-Zatsepin-Kuzmin cut-off of the spectrum, and we propose that they are originated from decays of superheavy long-living $CDM$-particles. 
These particles may have been  produced in the early Universe from our cosmic string after inflation and may constitute a considerable fraction of Cold Dark Matter. These $CDM$ particles are supersymmetric fermionic particles that can be produced by some cosmic string mechanism.
In some cases, induced isocurvature density fluctuations can leave an imprint in the anisotropy of cosmic microwave background radiation. The fermionic mass outside the string  is given by (\ref{mass2}): $M_F = \beta \, \eta$. We consider that the masses of these $CDM$ particles are of the order of $10^{13}$ GeV \cite{Kuzmin:1998kk} and they are in a range compatible  with  supersymmetric scales. We have a coupling parameter, $\mu = 10^{-3} $, and the cosmic string gauge field breaking parameter is of the order of $\eta = 10^{16}$ GeV, that corresponds to the energy scale at the end of the inflation. This parameter can work as a constraint for the coupling constant of the superpotential term (\ref{supot}).

\section{Fermionic Neutral Particles and Magnetic Moment Coupling}

Now, we pay attention to the neutral particles present in our model and we focus on their magnetic moment interaction. These particles, like the charged particles of the previous Section, can be considered as Dark Matter. 
According to astronomical observations, there is clear evidence for additional, non-luminous matter (or dark matter) in gravitational interactions; we however do not still understand their nature, for instance, their masses and other quantum numbers. 
Therefore, it is important to analyze their properties and other possible interaction mechanisms they may exhibit. These particles are a relic of the early Universe (for this 
reason, in many cases their masses are very heavy), but there are  alternative production scenarios, where very light particles can also act as Cold Dark Matter, as in the case of the axion \cite{Sikivie:2007qm}. In our framework, we have SUSY scales that give us huge masses. 
However, these particles are hard  be detected. It is possible to have some ways to do that. The particles considered here are non-charged, but can interact electromagnetically to some extent. For instance,
the neutron is a neutral particle with a significant magnetic moment. Thus, we wish to work on the possibility that dark matter has a small electromagnetic coupling via its magnetic moment and this moment is a by-product of LIV, as we shall see.

The Lagrangian that contains the non-charged fields is: 

\begin{equation}
\label{mass1}
{\cal L}_{NC} =
i\bar X \gamma^{\mu} \partial_{\mu} X  +
 i \bar \Lambda \gamma^{\mu} \partial_{\mu} \Lambda \, ,
\end{equation}
where the spinor $X$ is a partner of $H_{\mu}$ and $\Lambda$ is a partner of the $A_{\mu}$,

\begin{eqnarray}
\begin{array}{ll}
X(x) = \Big(\begin{array}{ll}\chi_a(x)\\ 
\bar \chi^{\dot a}(x) \end{array}\Big)
& \Lambda(x) = \sqrt{2} \Big(\begin{array}{ll}\lambda_a(x)\\
\bar \lambda^{\dot a}(x) \end{array}\Big)\, .
\end{array}
\end{eqnarray} 

We consider the expansion of the fermionic field  $\Psi$ in the $\Lambda $ and $X $ basis as
$\Psi = g \Lambda + g \rho  X$, where the background field is given by
\begin{equation}
 \Psi(x) = \Big(\begin{array}{ll}\psi_a(x)\\
\bar \psi^{\dot a}(x) \end{array}\Big)\, .
\end{equation} 

Adopting the basis for the physical gauge fields defined as in (\ref{gaugebasis}), we find that the neutral fields have an interesting coupling to the electromagnetic fields as below:

\begin{equation}
{\cal L}_I =   \Big(g \bar \Lambda\sigma^{\mu \nu} H_{\mu \nu} \Lambda + g \rho \bar X \sigma^{\mu \nu} F_{\mu \nu} X  \Big)\, ,\label{casher}
\end{equation}
where $g$ is a real constant
and the term responsible for the decay is
\begin{equation}
{\cal L}_{decay} =  g \bar \Lambda\sigma^{\mu \nu} F_{\mu \nu} X \, .\label{casher1}
\end{equation}

These fields present a mass term given by the coupling to the background density according to

\begin{equation}
{\cal L}^{M}_{NC} = - m (\bar X X + \bar \Lambda \Lambda)\, .\label{mass3}
\end{equation}
where $m = {8 \rho^2
|\Psi|^2 \over (1- \rho^2)^2 }$.
The Lagrangian (\ref{mass3}) has the form of a mass term, where $ | \Psi|^2 $ can be interpreted as the density. To understand the interaction Lagrangians (\ref{mass1}), (\ref{casher}) and (\ref{mass3}), let us analyze the equation of the motion for the fermionic field $ X$, given by:

\begin{equation}
\Big(i \gamma^{\mu} \partial_{\mu} + \rho \sigma_{\mu \nu} F^{\mu \nu} - M \Big) X =0,\label{equationcasher}
\end{equation}
where the  mass particle is $M = {m \over g \rho} $.
We have now a physical interpretation for the behavior of these neutral particles in connection with the magnetic moment. Aharonov and Casher proposed, in 1984, an experiment where they showed that there exists a phenomenon, in analogy with the Aharonov-Bohm effect, that involves the dynamics of a magnetic dipole moment 
in the presence of an external electric field \cite{Aharonov:1984xb}. We can show, with the help of the Lagrangian (\ref{casher}), that the cosmic string may be the source of the magnetic moment \cite{Bertone:2004pz} of the neutral supersymmetric massive particles that interact with the electric field, giving us an equation of motion as in (\ref{equationcasher}).

By analysing the electric and magnetic fields generated in our system, we find that the electric field outside the string is given by $E  =   - \frac{Q}{4\pi \epsilon_0 r}$  \,   for \, $v_z \neq 0$ and  $B_z =  - \frac{\mu_0 J}{4 \pi r}$ \, for \, $v_t \neq 0$.
This result shows that the breaking of Lorentz symmetry yields both electric charge and current associated to the magnetic flux related with the z-projection in the electric case and t-projection in the magnetic case of the violating background vector. The interesting point to be analyzed here is the fact that the parameters $\mu_0$ and $\epsilon_0$ are related with the Lorentz-symmetry breaking parameter $\rho$, that represents the fixed background.
Considering only the electric field and using the same procedure, we find that the magnetic moment in an external electric field gives us the Hamiltonian $H_{_{NR}} = (1/2M) \Big( {\vec p} -{\vec E }\times {\vec \mu}\Big)^2 -  \mu^2 E^2/M \, ,$
where, in the second term, there appears the correction induced by the electric 
dipole moment. The notation we used is ${ \mu} = 2 \rho =  { \kappa \,e  \over 2 M}$, which, for consistency, gives as $ {m\over g} = {\kappa e \over 4}$, where $ \kappa$ is the gyromagnetic ratio.


\section{General Conclusions and Remarks}

In this work, we have contemplated the possibility of formation of a cosmic
string configuration in a supersymmetric scenario where there is Lorentz-
symmetry violation. We have also considered its implications in a cosmological context.
As we have discussed, the astronomical observations provide compelling evidence for additional, 
non-luminous matter, or dark matter, and the most plausible theory that govern
these particles should be based on SUSY. On the other hand, there are evidences that
the high-energy events in our Universe can point to a LIV and there are
observations of the excess emission amplitude that gives us an agreement in 
temperature density fluctuation with the cosmic microwave background, showing that the
Universe may be different from what one has proposed until now 
\cite{Fixsen:2009xn,Seiffert:2009xs,Kogut:2009xv}.
It may happen that, in some era of the Universe,  LIV should not be discarded. With these
implications, it is very important to analyze the theoretical and experimental aspects
of this scenario. 

We show that, with our model, it is possible to have a discussion on    
the fermionic charged  supersymmetric massive particles.  These particles 
appear as SUSY partners in the same chiral scalar superfields that
accommodate the cosmic string scalar fields. Their masses are originated from
their Yukawa couplings, and so they are connected with the cosmic string breaking
scale, given by the scalar field vacuum expectation value, $\eta $, in the order
of $10^{16}$ GeV. In our approach, we use the Yukawa coupling constant ($\mu$) of the 
order of $10^3$ to give us particles with an energy compatible
with $10^{13}$ GeV. The parameter $\mu$ can be interpreted as the responsible for the 
fine tuning and can be adjusted with the experimental data.

We have chosen this mass scale  to take into account the experimental evidences
of the highly energetic cosmic rays, above the GZK cut-off of the spectrum,
that could be originated from decays of superheavy long-living
X-particles \cite{Kuzmin:1998kk}. These particles could be produced in the early
Universe from vacuum fluctuations during (and, in our case, after) the inflation,
when the cosmic string formation and the SUSY breaking took place.
There is another fine tuning parameter in our model, given by the LIV, that we have
denoted by $\rho $; it  parameterize the magnetic flux and has
non-trivial consequences on the analysis of the range of the gauge fields. 

Another contribution of this work is the study of the magnetic moment of the neutral dark matter
particles. This magnetic moment could be a way to detect them. The idea
is that these particles can present electromagnetic interactions to some extent.
We know of particles of the SM in a similar situation; for instance, the
neutron, that does not present electric charge, but has a significant magnetic
moment\cite{Gardner:2008yn}. In our model, we show that the Aharonov-Casher may occur and
an electric interaction with the magnetic moment of neutral particles takes place. This is 
a feature of the LIV consequences in a supersymemtric framework.

In our discussion, we may consider that our model can describe the axino. After a period of inflationary expansion, the Universe established a full thermal equilibrium at the temperature $T_R$. In the Large Hadron Colider (LHC) measurements, we can determine the temperature, $T_R$, in terms of the mass of the dark matter 
particles. Astrophysical and cosmological observations give us the determination of the relic density of the cold dark matter in the range of the $0,104 \pm 0,009$ \cite{Spergel:2006hy}.  These results may be adopted to impose constraints on the LIV parameters. This question is under consideration and we shall be reporting on it in a forthcoming paper (\cite{CA}).

{\bf Acknowledgments:}

CNF and JAH-N would like to express their gratitude to CNPq-Brasil for the invaluable financial support.


\begin{thebibliography}{200}

\bibitem{CPT} J.~S.~Schwinger,
  Phys.\ Rev.\  {\bf 82}, 914 (1951);
G. L\"uders and B. Zumino, 
Phys. Rev. {\bf 106}, 385, (1957): For experimental tests to see: R.~M.~Barnett {\it et al.}  [Particle Data Group],
  Phys.\ Rev.\  D {\bf 54}, 1 (1996);
B. Schwingenheuer et al., 
Phys. Rev. Lett. {\bf 74}, 4376 (1995); 
R.~Carosi {\it et al.}  [NA31 Collaboration],
  Phys.\ Lett.\  B {\bf 237}, 303 (1990).

\bibitem{Weinberg} S. Weinberg, The Quantum Theory of Fields, 
Vol. III, Cambridge University Press, Cambridge, 2000; 
J. Wess and J. Bagger, Supersymmetry and Supergravity, 
2nd ed., Princeton University Press, Princeton, 1992.

\bibitem{Colloday98} D.~Colladay and V.~A.~Kostelecky,
  Phys.\ Rev.\  D {\bf 55}, 6760 (1997)
  Phys.\ Rev.\  D {\bf 58}, 116002 (1998)

\bibitem{Kostelecky89} V.~A.~Kostelecky and S.~Samuel,
  Phys.\ Rev.\  D {\bf 40}, 1886 (1989);
 V.~A.~Kostelecky and R.~Potting,
  Nucl.\ Phys.\  B {\bf 359}, 545 (1991).

\bibitem{Kostelecky99} V. A. Kostelecky, ed, CPT and Lorentz Symmetry, 
World Scientific, Singapore, 1999.



\bibitem{Carroll:1989vb}
  S.~M.~Carroll, G.~B.~Field and R.~Jackiw,
  Phys.\ Rev.\  D {\bf 41}, 1231 (1990).



\bibitem{Stecker:1992wi}
  F.~W.~Stecker, O.~C.~de Jager and M.~H.~Salamon,
  Astrophys.\ J.\  {\bf 390}, L49 (1992).

\bibitem{Jacobson:2003bn}
  T.~A.~Jacobson, S.~Liberati, D.~Mattingly and F.~W.~Stecker,
  Phys.\ Rev.\ Lett.\  {\bf 93}, 021101 (2004)

\bibitem{Stecker:1998ib}
  F.~W.~Stecker and M.~H.~Salamon,
  Astrophys.\ J.\  {\bf 512}, 521 (1999)

\bibitem{Stecker:2004xm}
  F.~W.~Stecker and S.~T.~Scully,
  Astropart.\ Phys.\  {\bf 23}, 203 (2005)

\bibitem{Stecker:2004vm}
  F.~W.~Stecker,
  Int.\ J.\ Mod.\ Phys.\  A {\bf 20}, 3139 (2005)
  [arXiv:astro-ph/0409731].

\bibitem{Kostelecky95} V. A. Kostelecky, R. Potting,  
Phys. Lett. {\bf B 381}, 89 (1996);  
Phys Rev {\bf D 51}, 3923 (1996); 

\bibitem{Green87} Green M, Schwarz J. and Witten E, 
"Superstring Theory, vol 2 (Cambridge: Cambridge 
University Press), (1987).




\bibitem{Ferreira:2008zz} C.~N.~Ferreira, J.~A.~Helayel-Neto and W.~G.~Ney,
Phys.\ Rev.\  D {\bf 77}, 105028 (2008). 

\bibitem{Davis:1989gn} R.~L.~Davis and E.~P.~S.~Shellard,
Phys.\ Rev.\ Lett.\  {\bf 63}, 2021 (1989).

\bibitem{Majumdar:1999jd}
  P.~Majumdar and S.~SenGupta,
  Class.\ Quant.\ Grav.\  {\bf 16}, L89 (1999)
  [arXiv:gr-qc/9906027].


\bibitem{Maity:2004he}
  D.~Maity, P.~Majumdar and S.~SenGupta,
  JCAP {\bf 0406}, 005 (2004)
  [arXiv:hep-th/0401218].

\bibitem{Braga:2004ns}
  N.~R.~F.~Braga and C.~N.~Ferreira,
  JHEP {\bf 0503}, 039 (2005)
  [arXiv:hep-th/0410186].

\bibitem{Ferreira:2002mg}
  C.~N.~Ferreira, C.~F.~L.~Godinho and J.~A.~Helayel-Neto,
  New J.\ Phys.\  {\bf 6}, 58 (2004)
  [arXiv:hep-th/0205035].


\bibitem{Morris:1995wd}
  J.~R.~Morris,
  Phys.\ Rev.\  D {\bf 53}, 2078 (1996)
  [arXiv:hep-ph/9511293].


\bibitem{Davis:1997bs}
  S.~C.~Davis, A.~C.~Davis and M.~Trodden,
  Phys.\ Lett.\  B {\bf 405}, 257 (1997)
  [arXiv:hep-ph/9702360].



\bibitem{Ferreira:2000pi}
  C.~N.~Ferreira, M.~B.~D.~Porto and J.~A.~Helayel-Neto,
  Nucl.\ Phys.\  B {\bf 620}, 181 (2002)

\bibitem{NunesFerreira:2005if}
  C.~Nunes Ferreira, H.~Chavez and J.~A.~Helayel-Neto,
  PoS {\bf WC2004}, 036 (2004)
  [arXiv:hep-th/0501253].

\bibitem{Kibble:1980mv}
  T.~W.~B.~Kibble,
  Phys.\ Rept.\  {\bf 67}, 183 (1980).

\bibitem{Vilenkin:1981zs}
  A.~Vilenkin,
  Phys.\ Rev.\  D {\bf 23}, 852 (1981).

\bibitem{Hiscock:1985uc}
  W.~A.~Hiscock,
  Phys.\ Rev.\  D {\bf 31}, 3288 (1985).

\bibitem{Gott:1984ef}
  J.~R.~I.~Gott,
  Astrophys.\ J.\  {\bf 288}, 422 (1985).

\bibitem{Garfinkle:1985hr}
  D.~Garfinkle,
  Phys.\ Rev.\  D {\bf 32}, 1323 (1985).

\bibitem{Hindmarsh:1994re}
  M.~B.~Hindmarsh and T.~W.~B.~Kibble,
  Rept.\ Prog.\ Phys.\  {\bf 58}, 477 (1995)
  [arXiv:hep-ph/9411342].

\bibitem{Vilenkin1} A.Vilenkin and 
E.P.S.Shellard, {\it Cosmic Strings and
other Topological Defects} 
(Cambridge University Press, 1994).

\bibitem{Kibble:1976sj}
  T.~W.~B.~Kibble,
  J.\ Phys.\ A  {\bf 9}, 1387 (1976).

\bibitem{Sato:1986pd}
  Y.~Sato,
  Prog.\ Theor.\ Phys.\  {\bf 75}, 914 (1986).


\bibitem{Turok:1985tt}
  N.~Turok and R.~H.~Brandenberger,
  Phys.\ Rev.\  D {\bf 33}, 2175 (1986).


\bibitem{Kibble:2004hq}
  T.~W.~B.~Kibble,
  arXiv:astro-ph/0410073.

\bibitem{Polchinski:2004ia}
  J.~Polchinski,
  arXiv:hep-th/0412244.

\bibitem{Bhattacharjee:1989vu}
  P.~Bhattacharjee,
  Phys.\ Rev.\  D {\bf 40}, 3968 (1989).


\bibitem{Bhattacharjee:1991zm}
  P.~Bhattacharjee, C.~T.~Hill and D.~N.~Schramm,
  Phys.\ Rev.\ Lett.\  {\bf 69}, 567 (1992).


\bibitem{MacGibbon:1989kk}
  J.~H.~MacGibbon and R.~H.~Brandenberger,
  Nucl.\ Phys.\  B {\bf 331}, 153 (1990).


\bibitem{Kuzmin:1998kk}
  V.~Kuzmin and I.~Tkachev,
  Phys.\ Rev.\  D {\bf 59}, 123006 (1999)
  [arXiv:hep-ph/9809547].


\cite{Bezerra:2004qv}
\bibitem{Bezerra:2004qv}
  V.~B.~Bezerra, C.~N.~Ferreira and J.~A.~Helayel-Neto,
  Phys.\ Rev.\  D {\bf 71}, 044018 (2005)
  [arXiv:hep-th/0405181].

\bibitem{Belich:2003fa}
  H.~Belich, J.~L.~Boldo, L.~P.~Colatto, J.~A.~Helayel-Neto and A.~L.~M.~Nogueira,
  Phys.\ Rev.\  D {\bf 68}, 065030 (2003)
  [arXiv:hep-th/0304166].



\bibitem{Witten:1986qx}
  E.~Witten,
{\it  In *Chicago 1986, Proceedings, Relativistic Astrophysics* 606-620. }


\bibitem{Foot:1994vd}
R.~Foot, X.~G.~He, H.~Lew and R.~R.~Volkas,
Phys.\ Rev.\  D {\bf 50}, 4571 (1994)


\bibitem{Kuhfuss:1986rb}
  R.~Kuhfuss and J.~Nitsch,
  Gen.\ Rel.\ Grav.\  {\bf 18}, 1207 (1986).


\bibitem{Sikivie:2007qm}
  P.~Sikivie, D.~B.~Tanner and K.~van Bibber,
  Phys.\ Rev.\ Lett.\  {\bf 98}, 172002 (2007)
  [arXiv:hep-ph/0701198].


\bibitem{Aharonov:1984xb}
  Y.~Aharonov and A.~Casher,
  Phys.\ Rev.\ Lett.\  {\bf 53} (1984) 319.


\bibitem{Bertone:2004pz}
  G.~Bertone, D.~Hooper and J.~Silk,
  Phys.\ Rept.\  {\bf 405}, 279 (2005)
  [arXiv:hep-ph/0404175].


\bibitem{Fixsen:2009xn}
  D.~J.~Fixsen {\it et al.},
  arXiv:0901.0555 [astro-ph.CO].

\bibitem{Seiffert:2009xs}
  M.~Seiffert {\it et al.},
  arXiv:0901.0559 [astro-ph.CO].

\bibitem{Kogut:2009xv}
  A.~Kogut {\it et al.},
  arXiv:0901.0562 [astro-ph.GA].

\bibitem{Gardner:2008yn}
  S.~Gardner,
  Phys.\ Rev.\  D {\bf 79}, 055007 (2009)
  [arXiv:0811.0967 [hep-ph]].




\bibitem{Spergel:2006hy}
  D.~N.~Spergel {\it et al.}  [WMAP Collaboration],
  Astrophys.\ J.\ Suppl.\  {\bf 170}, 377 (2007).

\bibitem{Kim:1984yn}
  J.~E.~Kim, A.~Masiero and D.~V.~Nanopoulos,
  Phys.\ Lett.\  B {\bf 139}, 346 (1984).

\bibitem{CA} C. N. Ferreira, L. A. S. Nunes, C. A. Almeida and J. A. Helayel-Neto, work in progress.


\end{thebibliography}
\end{document}